\documentclass[11pt]{article}
 
\usepackage{amsmath}
\usepackage{amssymb}
\usepackage{amsbsy}
\usepackage{epic}
\usepackage{eepic}
\usepackage{times}
\usepackage{eucal}
\newtheorem{theorem}{Theorem}[section]
\newtheorem{lemma}[theorem]{Lemma}
 
\newtheorem{fact}[theorem]{Fact}
\newcommand{\ket}[1]{\left \vert #1 \right \rangle}

\newcommand{\braket}[2]{\left \langle #1 \left \vert #2 \right \rangle\right .}
\newcommand{\density}[2]{\left\vert #1 \left\rangle\right\langle #2 \right\vert} 
\newcommand{\norm}[1]{\left\lVert #1 \right\rVert}
\newcommand{\rr}{\mbox{$\mathbb R$}} 
\newcommand{\cc}{\mbox{$\mathbb C$}}
\newcommand{\bsym}{\boldsymbol} 
\newcommand{\ov}{\overline} 

\newcommand{\vep}{\varepsilon} 
\newcommand{\ft}{\footnotesize}
 
\newcommand{\qed}{\mbox{\rule{1.6mm}{4.3mm}}}

\title{Quantum Formulas: a Lower Bound and Simulation
\thanks{This work was supported in part by grants from the Revolutionary Computing
group at JPL (contract \#961360), and from the DARPA Ultra program
(subcontract from Purdue University \#530--1415--01).}} 

\author{Vwani P. Roychowdhury\thanks{Electrical Engineering Department, UCLA, 
Los Angeles, CA 90095 ({\tt vwani@ee.ucla.edu}).} \and
Farrokh Vatan\thanks{Electrical Engineering Department, UCLA, 
Los Angeles, CA 90095 ({\tt vatan@ee.ucla.edu}). Present address: 
Jet Propulsion Laboratory, California Institute of Technology,
4800 Oak Grove Drive Pasadena, CA 91109 ({\tt Farrokh.Vatan@jpl.nasa.gov}).}}

\date{ }

\begin{document} 

\maketitle 

\begin{abstract} 
We show that Nechiporuk's method \cite{wegener} for proving
lower bounds for Boolean formulas can be extended to the quantum case.
This leads to an $\Omega(n^2/\log^2 n)$ lower bound for quantum formulas
computing an explicit function. The only known previous explicit 
lower bound for quantum formulas \cite{yao} states that the majority function 
does not have a linear--size quantum formula. 
We also show that quantum formulas can be simulated
by Boolean circuits of almost the same size.

{\bf Key words.}
quantum formula, lower bound, mixed state, density matrix

{\bf AMS subject classification:} 81P68, 68Q10, 68Q05, 03D10

\begin{center}
To appear in {\em SIAM Journal on Computing}
\end{center}
\end{abstract} 

\section{Introduction} 

Computational devices based on quantum physics have attracted much
attention lately, and quantum algorithms that perform much faster than
their classical counterparts have been developed
\cite{grover,shor,simon}. To provide a systematic study of the
computational power of quantum devices, models similar to those for
classical computational devices have been proposed. Deutsch
\cite{deutsch85} formulated the notion of quantum Turing machine. This
approach was further developed by Bernstein and Vazirani \cite{vazirani},
and the concept of an efficient universal quantum Turing machine was
introduced. As in the case of classical Boolean computation, there is
also a quantum model of computation based on circuits (or networks). Yao
\cite{yao} proved that the quantum circuit model, first introduced by
Deutsch \cite{deutsch89}, is equivalent to the quantum Turing machine model. 

Since every Boolean circuit can be simulated by a quantum circuit, with
at most a polynomial factor increase in its size, any nontrivial lower bound 
for quantum circuits could have far reaching consequences. In classical
Boolean circuit theory, all nontrivial lower bounds are for proper
subclasses of Boolean circuits such as monotone circuits, formulas,
bounded-depth circuits, etc. In the quantum case also it seems that 
the only hope to prove nontrivial lower bounds is for proper subclasses of
quantum circuits. So far the only such known lower bound has been derived by Yao
\cite{yao} for quantum formulas.\footnote{There are exponential lower
bounds on the time of quantum computation for the black--box model 
(see, e.g., \cite{beals}), but
they do not apply to the size of quantum circuits.} The quantum formula is a
straightforward generalization of the classical Boolean formula: in both
cases, the graph of the circuit is a tree. Yao has proved that the
quantum formula size of the majority function $\mbox{MAJ}_n$ is not
linear\footnote{The value of $\mbox{MAJ}_n(x_1,\ldots,x_n)$ is $1$ if at
least $\lceil n/2\rceil$ of inputs are 1.}; i.e., if $L(\mbox{MAJ}_n)$ 
denotes the minimum quantum formula size of $\mbox{MAJ}_n$ then
$\lim_{n\longrightarrow\infty}L(\mbox{MAJ}_n)/n=\infty$. This bound is
derived from a bound on the quantum communication complexity of Boolean
functions. 

In this paper, we prove an almost quadratic lower bound for quantum
formula size. The key step in the derivation of this lower bound is the
extension of Nechiporuk's method to quantum formulas; for a detailed
discussion of Nechiporuk's method in the Boolean setting see
\cite{dunne,wegener}. Nechiporuk's method has been used in several
different areas of Boolean complexity (e.g., see \cite{dunne} for details). It
has also been applied to models where the gates do not take on binary or
discrete values, but the input/output map still corresponds to a Boolean
function. For example, in \cite{turan} this method has been used to get
a lower bound for arithmetic and threshold formulas. The challenging
part of this method is a step that we shall refer to as ``path
squeezing'' (see \S\ref{lowerbound} for the exact meaning of it). 
Although in the
case of Boolean gates, this part can be solved easily, in the case of
analog circuits it is far from obvious (see \cite{turan}). For the
quantum formulas ``path squeezing'' becomes even more complicated,
because here we should take care of any {\em quantum entanglement}\/ and
interference phenomena. We show that it is still possible to squeeze a
path with arbitrary number of constant inputs to a path
with a fixed number of inputs. This leads to a lower bound of $\Omega(n^2/\log^2 n)$ 
on the size of quantum formulas computing a class of explicit functions. 
For example, we get such a bound for the Element Distinctness function $\mbox{ED}_n$.
The input of $\mbox{ED}_n$, for $n=2\ell\log\ell$,
 is of the form $(z_1,\ldots,z_\ell)$, where each $z_j$ is
a string of $2\log\ell$ bits. Then $\mbox{ED}_n(z_1,\ldots,z_\ell)=1$ if and only
if all these strings are pair wise distinct.

In the end of the paper we compare the powers of quantum formulas 
and Boolean {\em circuits}. Surprisingly, in some sense quantum formulas are 
not more powerful than Boolean circuits. Any quantum formula
of size $s$ and depth $d$ can be approximated by a Boolean circuit of size
$O(s\log s\log\log s)$ and depth $O(d\log\log s)$. Similar results
are not known, and most probably are not true, for quantum circuits and other models
which are depending on real number parameters (like arithmetic circuits \cite{turan}).
The key idea for this simulation is that the computation of a quantum formula on
an input (which is a pure state in the Hilbert space) can be described as performing
a sequence of unitary operations on $4\times 4$ density matrices of mixed states.

In this paper we use the notation $|\cdot|$ for two different purposes. When $\alpha$ 
is a complex number, $|\alpha|$ denotes the absolute value of $\alpha$; i.e.,
$|\alpha|=\sqrt{\alpha\cdot\alpha ^*}$. While if $X$ is a set then $|X|$
denotes the cardinality of $X$.

\section{Preliminaries} 

A {\em quantum circuit}\/ is defined as a straightforward generalization of
acyclic classical (Boolean) circuit (see \cite{deutsch89}). 
For constructing a quantum circuit, we begin with a {\em basis}\/ of
quantum gates as elementary gates.
Each elementary gate $g$ with $d$ inputs represents a unitary
operation $U_g\in\mbox{\bf U}(2^d)$, where $\mbox{\bf U}(m)$ denotes the
group of $m\times m$ unitary complex matrices.
The gates are interconnected by quantum ``wires''. Each wire
represents a quantum bit, {\em qubit}, which is a 2--state quantum
system represented by a unit vector in $\cc^2$. Let
$\{\ket{0},\ket{1}\}$ be the standard orthonormal basis of $\cc^2$. The
$\ket{0}$ and $\ket{1}$ values of a qubit correspond to the classical
Boolean $0$ and $1$ values, but a qubit can also be in a superposition
of the form $\alpha\ket{0}+\beta\ket{1}$, where $\alpha,\beta\in\cc$ and
$|\alpha|^2+|\beta|^2=1$. 
Note that the output of such gate, in general,
is not a tensor product of its inputs, but an {\em entangled state}; e.g., a
state like $\frac{1}{\sqrt{2}}\ket{00}+\frac{1}{\sqrt{2}}\ket{11}$ which can not
be written as a tensor product.

If the circuit has $m$ inputs, then for each $d$--input gate $g$, the unitary
operation $U_g\in\mbox{\bf U}(2^d)$ can be considered in a natural way as an
operator in $\mbox{\bf U}(2^m)$ by acting as the identity operator on the
other $(m-d)$ qubits. Hence, a quantum circuit with $m$ inputs computes a
unitary operator in $\mbox{\bf U}(2^m)$, which is the product of successive
unitary operators defined by successive gates. 

The {\em size}\/ of a quantum circuit $C$, denoted by $\mathsf{size}(C)$, 
is the number of gates occurring in $C$. The {\em depth}\/ of $C$, denoted
by $\mathsf{depth}(C)$, is the length of the longest path in $C$ from an input
to an output gate.

In this paper, we consider quantum circuits that compute Boolean
functions. Consider a quantum circuit $C$ with $m$ inputs. Suppose that
$C$ computes the unitary operator $U_C\in\mbox{\bf U}(2^m)$. We say $C$
computes the Boolean function $f\colon\{0,1\}^n\longrightarrow \{0,1\}$
if the following holds. The inputs are labeled by the variables
$x_1,x_2,\ldots, x_n$ or the constants $\ket{0}$ or $\ket{1}$ (different
inputs may be labeled by the same variable $x_j$). We consider one of
the output wires, say the first one, as the output of the
circuit. To compute the value of the circuit at
$(a_1,\ldots,a_n)\in\{0,1\}^n$, let the value of each
input wire with label $x_j$ be $\ket{a_j}$. These inputs, along
with the constant inputs to the circuit, define a unit vector $\ket{\alpha}$ 
in $\cc^{2^m}$. In fact this
vector is a standard basis vector of the following form (up to some
repetitions and a permutation) 
\[\ket{\alpha}=\ket{a_1}\otimes\cdots\otimes\ket{a_n}\otimes
    \ket{0}\otimes\cdots\otimes\ket{1}. \] 
The action of the circuit $C$ on the
input $\ket{\alpha}$ is the same as $U_C(\ket{\alpha})$. Note that since
$U_C$ is unitary, $\norm{U_C(\ket{\alpha})}=1$. We decompose the vector
$U_C(\ket{\alpha})\in\cc^{2^m}$ with respect to the output qubit. Let
the result be 
\[ U_C(\ket{\alpha}) = \ket{0}\otimes\ket{A_{0,\alpha}}+
          \ket{1}\otimes \ket{A_{1,\alpha}} .\] 
Then we define the {\em probability}\/ that $C$ outputs 1 (on the input $\alpha$) 
as $p_\alpha=\norm{\ket{A_{1,\alpha}}}^2$, i.e., the square of the length of
$\ket{A_{1,\alpha}}\in\cc^{2^{m-1}}$. Finally, we say that the quantum circuit
$C$ computes the Boolean function $f$ if for every $\alpha\in\{0,1\}^n$,
if $f(\alpha)=1$ then $p_\alpha > 2/3$ and if $f(\alpha)=0$ then
$p_\alpha < 1/3$. 

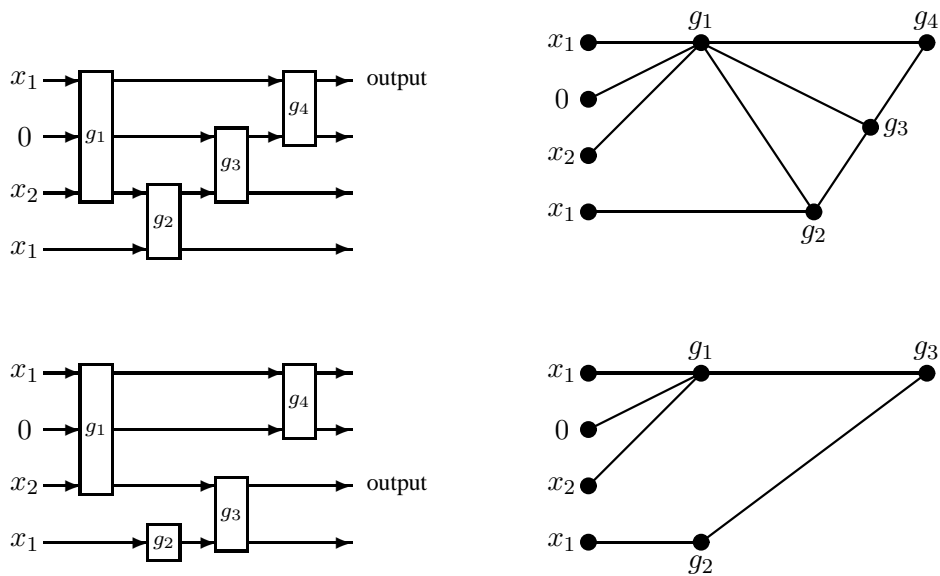
\begin{figure} 
\thicklines
\begin{center} 
\unitlength=.5mm 
\begin{tabular}{cp{8mm}c}
\begin{picture}(110,51)(0,2) \put(0,0){\makebox(0,0){$x_1$}}
\put(0,15){\makebox(0,0){$x_2$}} \put(0,30){\makebox(0,0){$0$}}
\put(0,45){\makebox(0,0){$x_1$}} \put(5,15){\vector(1,0){10}}
\put(5,30){\vector(1,0){10}} \put(5,45){\vector(1,0){10}}
\put(19,30){\makebox(0,0){\framebox(8,34){$\scriptstyle g_1$}}}
\put(5,0){\vector(1,0){28}} \put(23,15){\vector(1,0){10}}
\put(37,7.5){\makebox(0,0){\framebox(8,19){$\scriptstyle g_2$}}}
\put(41,15){\vector(1,0){10}} \put(23,30){\vector(1,0){28}}
\put(55,22.5){\makebox(0,0){\framebox(8,19){$\scriptstyle g_3$}}}
\put(41,0){\vector(1,0){46}} \put(59,15){\vector(1,0){28}}
\put(59,30){\vector(1,0){10}} \put(23,45){\vector(1,0){46}}
\put(73,37.5){\makebox(0,0){\framebox(8,19){$\scriptstyle g_4$}}}
\put(77,30){\vector(1,0){10}} \put(77,45){\vector(1,0){10}}
\put(99,45){\makebox(0,0){\ft output}} \end{picture} 

& & 

\begin{picture}(100,61)(0,-8) 
\put(0,0){\makebox(0,0){$x_1$}}
\put(0,15){\makebox(0,0){$x_2$}} \put(0,30){\makebox(0,0){$0$}}
\put(0,45){\makebox(0,0){$x_1$}} \put(7,0){\circle*{4}}
\put(7,15){\circle*{4}} \put(7,30){\circle*{4}} \put(7,45){\circle*{4}}
\put(7,45){\line(1,0){90}} \put(37,45){\circle*{4}}
\put(37,51){\makebox(0,0){$g_1$}} \put(97,45){\circle*{4}}
\put(97,51){\makebox(0,0){$g_4$}} \put(7,30){\line(2,1){30}}
\put(7,15){\line(1,1){30}} \put(7,0){\line(1,0){60}}
\put(67,0){\circle*{4}} \put(67,-6){\makebox(0,0){$g_2$}}
\put(67,0){\line(2,3){30}} \put(67,0){\line(-2,3){30}}
\put(82,22.5){\circle*{4}} \put(89,22.5){\makebox(0,0){$g_3$}}
\put(82,22.5){\line(-2,1){45}} 

\end{picture} 

\\ \vspace{10mm} & \\ 

\begin{picture}(110,50)(0,-5) 
\put(0,0){\makebox(0,0){$x_1$}}
\put(0,15){\makebox(0,0){$x_2$}} \put(0,30){\makebox(0,0){$0$}}
\put(0,45){\makebox(0,0){$x_1$}} \put(5,15){\vector(1,0){10}}
\put(5,30){\vector(1,0){10}} \put(5,45){\vector(1,0){10}}
\put(19,30){\makebox(0,0){\framebox(8,34){$\scriptstyle g_1$}}}
\put(5,0){\vector(1,0){28}} 
\put(37,0){\makebox(0,0){\framebox(8,9){$\scriptstyle g_2$}}} 
\put(41,0){\vector(1,0){10}} \put(23,15){\vector(1,0){28}}
\put(55,7.5){\makebox(0,0){\framebox(8,19){$\scriptstyle g_3$}}}
\put(59,0){\vector(1,0){28}} \put(59,15){\vector(1,0){28}}
\put(23,30){\vector(1,0){46}} \put(23,45){\vector(1,0){46}}
\put(73,37.5){\makebox(0,0){\framebox(8,19){$\scriptstyle g_4$}}}
\put(77,30){\vector(1,0){10}} \put(77,45){\vector(1,0){10}}
\put(99,15){\makebox(0,0){\ft output}} \end{picture} 

& & 

\begin{picture}(100,50)(0,-5) \put(0,0){\makebox(0,0){$x_1$}}
\put(0,15){\makebox(0,0){$x_2$}} \put(0,30){\makebox(0,0){$0$}}
\put(0,45){\makebox(0,0){$x_1$}} \put(7,0){\circle*{4}}
\put(7,15){\circle*{4}} \put(7,30){\circle*{4}} \put(7,45){\circle*{4}}
\put(7,45){\line(1,0){90}} \put(37,45){\circle*{4}}
\put(37,51){\makebox(0,0){$g_1$}} \put(97,45){\circle*{4}}
\put(97,51){\makebox(0,0){$g_3$}} \put(7,30){\line(2,1){30}}
\put(7,15){\line(1,1){30}} \put(7,0){\line(1,0){30}}
\put(37,0){\circle*{4}} \put(37,-6){\makebox(0,0){$g_2$}}
\put(37,0){\line(4,3){60}} 

\end{picture}\end{tabular}\end{center}
\caption{Quantum circuits and their computation graphs; the top circuit
is not a formula while the bottom one is a formula.} 
\label{fig1} 
\end{figure}

Following Yao \cite{yao}, we define quantum formulas as a subclass of
quantum circuits. A quantum circuit $C$ is a {\em formula}\/ if for every
input there is a unique path that connects it to the output qubit. To
make this definition more clear we define the {\em computation graph}\/ of
$C$, denoted by $G_C$. The nodes of $G_C$ correspond to a subset of the gates of
$C$. We start with the output gate of $C$, i.e., the gate which provides
the output qubit, and let it be a node of $G_C$. Once a node $v$ belongs
to $G_C$ then all gates in $C$ that provide inputs to $v$ are considered
as adjacent nodes of $v$ in $G_C$. Then $C$ is a formula if the graph
$G_C$ is a tree. Figure~\ref{fig1} provides examples of quantum circuits of both
kinds, i.e., circuits that are also quantum formulas, and circuits that
are not formulas.

All circuits that we consider are over some fixed  quantum
basis. The lower bound does not depend on the basis; the only condition
is that the number of inputs (and so the number of outputs) of each gate
be bounded by some fixed constant number (this condition is usually
considered as part of the definition of a quantum basis). For example,
this basis can be the set of all 2--input 2--output quantum gates, and as
as it is shown in \cite{barenco-etal}, this basis is universal for computation 
with quantum circuits. 

It is well--known that any Boolean circuit can be efficiently simulated by a 
quantum circuit over a universal basis.
Indeed, for this purpose, the 3--bit {\em Toffoli gate}\/ is enough
(see, e.g., \cite{bennett,levine}).
Similarly, any Boolean formula can be efficiently simulated by a quantum formula 
using only Toffoli gate or a basis universal for classical computation. 
In the special case,
from \cite{valiant} it follows that there is a polynomial--size log--depth
quantum {\em formula}\/ computing the majority function $\mathrm{MAJ}_n$. 
This fact implies that for quantum formulas over reasonable bases (i.e., 
universal for classical computation) the threshold probability of
correct answer ($\frac{2}{3}$ in the above definition) can be efficiently 
boosted to a number arbitrarily close to one.

For our proof we also need a Shannon--type result for quantum circuits.
Knill \cite{knill} has proved several theorems about the quantum circuit
complexity of almost all Boolean functions. We will use the following
theorem. 

\begin{theorem} [{\rm \cite{knill}}] 
The number of different $n$--variable
Boolean functions that can be computed by size $N$ quantum circuits ($n\leq
N$) with $d$--input $d$--output elementary gates is at most $2^{cN\log
N}$, where $c$ depends only on $d$. 
\label{knill} 
\end{theorem} 

For the sake of completeness, in Appendix we have provided a proof for a
slightly weaker bound. Our approach is different from that in \cite{knill} and it 
seems it is shorter and simpler than the proof in \cite{knill}. Although the 
bound that we get is a little weaker than the bound provided by the above theorem 
(it is of the form $2^{O(nN)}$), our bound results in the same bound of
Theorem~\ref{knill} if $\log(N)=\Omega(n)$ which is true for almost all 
Boolean functions. Thus our result provides the same bound for the complexity 
of almost all functions and it is sufficient for the bound we get in this paper.

We also need to consider general orthonormal bases in the space $\cc^{2^n}$
other than the standard basis. In the context of quantum physics, we
identify the Hilbert space $\cc^{2^n}$ as the tensor product space
$\bigotimes_{j=1}^n \cc^2$, and the standard basis consists of the
vectors 
\[\ket{c_1}\otimes\cdots\otimes\ket{c_n}=\ket{c_1\cdots
c_n},\ c_j\in\{0,1\} . \] 



\begin{fact} 
Let $\ket{A_j}\in\cc^{2^k}$ and $\ket{B_\ell}\in\cc^{2^m}$ be
unit vectors (for $j$ and $\ell$ in some index sets). If $\ket{A_j}$ are
pair wise orthogonal and $\ket{B_\ell}$ are pair wise orthogonal then the
family \[ \left\{ \ket{A_j}\otimes\ket{B_{\ell}}\in\cc^{2^{k+m}} \colon
j,\ell\right\} \] is an orthonormal set. 
\label{basis} 
\end{fact} 

The following lemma, although seemingly obvious, is crucial for the ``path squeezing'' 
technique in the proof of the lower bound.

\begin{figure} \thicklines\begin{center} \unitlength=.4mm 

\begin{picture}(286,66)(-16,-3) 
\put(-10,10){\makebox(0,0){$Q_2$}}
\put(-10,50){\makebox(0,0){$Q_1$}} \put(0,0){\line(1,0){10}}
\put(0,10){\line(1,0){10}} \put(0,20){\line(1,0){30}}
\put(0,40){\line(1,0){10}} \put(0,50){\line(1,0){10}}
\put(0,60){\line(1,0){10}} \put(15,50){\makebox(0,0){\framebox(10,24){
}}} \put(15,5){\makebox(0,0){\framebox(10,14){ }}}
\put(20,0){\line(1,0){30}} \put(20,10){\line(1,0){10}}
\put(20,40){\line(1,0){10}} \put(20,50){\line(1,0){30}}
\put(20,60){\line(1,0){10}} \put(35,15){\makebox(0,0){\framebox(10,14){
}}} \put(35,40){\makebox(0,0){\framebox(10,6){ }}}
\put(35,60){\makebox(0,0){\framebox(10,6){ }}}
\put(40,10){\line(1,0){10}} \put(40,20){\line(1,0){10}}
\put(40,40){\line(1,0){30}} \put(40,60){\line(1,0){10}}
\put(55,10){\makebox(0,0){\framebox(10,24){ }}}
\put(55,55){\makebox(0,0){\framebox(10,14){ }}}
\put(60,0){\line(1,0){10}} \put(60,10){\line(1,0){10}}
\put(60,20){\line(1,0){10}} \put(60,50){\line(1,0){10}}
\put(60,60){\line(1,0){10}} 

\put(85,30){\makebox(0,0){$=$}} 

\put(100,0){\line(1,0){40}} \put(100,10){\line(1,0){40}}
\put(100,20){\line(1,0){40}} \put(100,40){\line(1,0){10}}
\put(100,50){\line(1,0){10}} \put(100,60){\line(1,0){10}}
\put(120,50){\makebox(0,0){\framebox(20,24){$C_1$}}}
\put(150,10){\makebox(0,0){\framebox(20,24){$C_2$}}}
\put(160,0){\line(1,0){10}} \put(160,10){\line(1,0){10}}
\put(160,20){\line(1,0){10}} \put(130,40){\line(1,0){40}}
\put(130,50){\line(1,0){40}} \put(130,60){\line(1,0){40}} 

\put(185,30){\makebox(0,0){$=$}} 

\put(200,0){\line(1,0){10}} \put(200,10){\line(1,0){10}}
\put(200,20){\line(1,0){10}} \put(200,40){\line(1,0){40}}
\put(200,50){\line(1,0){40}} \put(200,60){\line(1,0){40}}
\put(250,50){\makebox(0,0){\framebox(20,24){$C_1$}}}
\put(220,10){\makebox(0,0){\framebox(20,24){$C_2$}}}
\put(230,0){\line(1,0){40}} \put(230,10){\line(1,0){40}}
\put(230,20){\line(1,0){40}} \put(260,40){\line(1,0){10}}
\put(260,50){\line(1,0){10}} \put(260,60){\line(1,0){10}} 

\end{picture} \end{center} 
\caption{Decomposition of a quantum subcircuit acting on disjoint sets of
qubits (Lemma~\protect\ref{composition} (a)).} 
\label{fig2}
\end{figure}
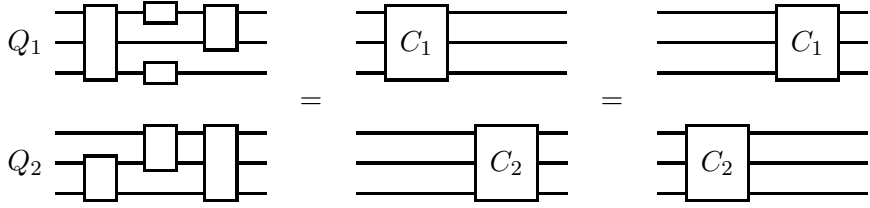

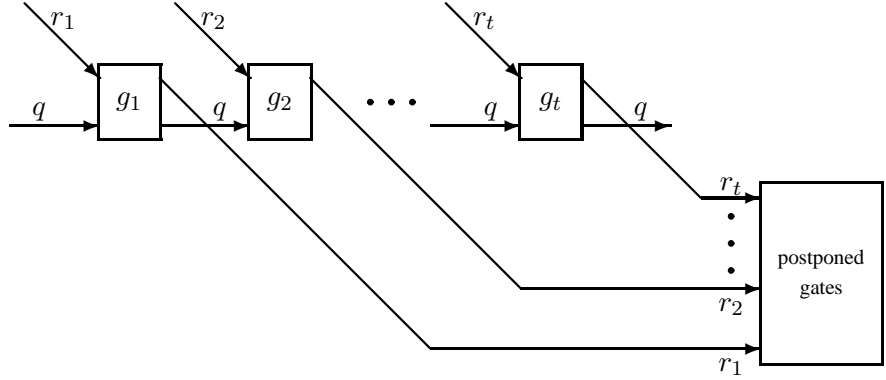
\begin{figure} 
\thicklines 
\begin{center} 
\unitlength=.4mm 
\begin{picture}(300,120)(0,-80) 
\put(0,0){\vector(1,0){30}}\put(30,16){\line(-1,1){25}}
\put(40,8){\makebox(0,0){\framebox(20,24){$g_1$}}}
\put(10,5){\makebox(0,0){$q$}}
\put(30,16){\vector(1,-1){0}}\put(18,35){\makebox(0,0){$r_1$}}

\put(50,0){\vector(1,0){30}}\put(80,16){\line(-1,1){25}}
\put(90,8){\makebox(0,0){\framebox(20,24){$g_2$}}}
\put(70,5){\makebox(0,0){$q$}}
\put(80,16){\vector(1,-1){0}}\put(68,35){\makebox(0,0){$r_2$}}
\put(50,16){\line(1,-1){90}}\put(140,-74){\vector(1,0){110}}
\put(100,16){\line(1,-1){70}}\put(170,-54){\vector(1,0){80}}

\put(140,0){\vector(1,0){30}}\put(170,16){\line(-1,1){25}}
\put(180,8){\makebox(0,0){\framebox(20,24){$g_t$}}}
\put(160,5){\makebox(0,0){$q$}}
\put(170,16){\vector(1,-1){0}}\put(158,35){\makebox(0,0){$r_t$}}
\put(190,0){\vector(1,0){30}}\put(210,5){\makebox(0,0){$q$}}
\put(190,16){\line(1,-1){40}}\put(230,-24){\vector(1,0){20}}
\put(270,-49){\makebox(0,0){\framebox(40,60)
{\scriptsize\begin{tabular}{c} postponed \\ gates\end{tabular}}}}

\put(120,8){\circle*{1.5}}\put(127,8){\circle*{1.5}}\put(134,8){\circle*{1.5}}
\put(240,-30){\circle*{1.5}}\put(240,-39){\circle*{1.5}}\put(240,-48){\circle*{1.5}}
\put(240,-80){\makebox(0,0){$r_1$}}\put(240,-60){\makebox(0,0){$r_2$}}
\put(240,-20){\makebox(0,0){$r_t$}}

\end{picture} \end{center} 
\caption{Postponing the gates (Lemma~\protect\ref{composition} (b)).} 
\label{fig3} 
\end{figure}

\begin{lemma} (a) Suppose that $C$ is a subcircuit of a quantum circuit. Let 
the inputs of $C$ be divided into two disjoint sets of qubits $Q_1$ and $Q_2$. 
Suppose that each gate 
of $C$ either acts only on qubits from $Q_1$ or only on qubits from $Q_2$. Then 
there are subcircuits $C_1$ and $C_2$ such that $C_j$ acts only on qubits from
$Q_j$ and the operation of $C$ is the composition of operations of $C_1$
and $C_2$ no matter in which order they act; i.e., $C=C_1\circ
C_2=C_2\circ C_1$. So the subcircuit $C$ can be substituted
by $C_1$ and $C_2$ (see Figure~{\em \ref{fig2}}).

(b) Let $C$ be a subcircuit of a quantum circuit with distinct input qubits $q$ 
and $r_1,\ldots,r_t$. Suppose that only $t$ gates $g_1,\ldots,g_t$ in 
$C$ act on $q$. Moreover, suppose that each $g_j$ acts 
on $q$ and $r_j$. Then, w.l.o.g., we can assume that each qubit $r_j$ after entering 
the gate $g_j$ will not interact with any other qubit until the gate $g_t$ is 
performed (see Figure~{\em \ref{fig3}}).
\label{composition} 
\end{lemma}

{\bf Proof.}
Part (a) is based on the following simple
observation. If $M\in\mbox{\bf U}(2^m)$ and $N\in\mbox{\bf U}(2^n)$ then
\begin{align*}
M\otimes N &= (M\otimes I_n)\circ(I_m\otimes N) \\
           &= (I_m\otimes N)\circ(M\otimes I_n), 
\end{align*}
where $I_t$ is the identity map in $\mbox{\bf U}(2^t)$.
Note that the inputs of the subcircuit $C$ may be in an entangled state; but
to see that the equality $C=C_1\circ C_2=C_2\circ C_1$ holds, it is enough 
to check this equality for the standard basis and extend it to the whole 
space by linearity.

\begin{figure}
\thicklines
\begin{center} 
\unitlength=.37mm 
\begin{picture}(330,65)(0,-35)
\put(0,0){\vector(1,0){20}}\put(20,16){\line(-1,1){15}}
\put(30,8){\makebox(0,0){\framebox(20,24){$h_1$}}}
\put(20,16){\vector(1,-1){0}}
\put(40,16){\vector(1,-1){30}}\put(50,-30){\vector(1,0){20}}
\put(80,-22){\makebox(0,0){\framebox(20,24){$h_2$}}}
\put(40,0){\vector(1,0){70}}\put(110,16){\line(-1,1){15}}
\put(120,8){\makebox(0,0){\framebox(20,24){$h_3$}}}
\put(110,16){\vector(1,-1){0}}
\put(130,0){\vector(1,0){20}}\put(130,16){\vector(1,0){20}}
\put(90,-30){\vector(1,0){20}}\put(90,-14){\vector(1,0){20}}

\put(165,-4){\makebox(0,0){$=$}}

\put(180,0){\vector(1,0){20}}\put(200,16){\line(-1,1){15}}
\put(210,8){\makebox(0,0){\framebox(20,24){$h_1$}}}
\put(200,16){\vector(1,-1){0}} 
\put(220,0){\vector(1,0){30}}\put(250,16){\line(-1,1){15}}
\put(260,8){\makebox(0,0){\framebox(20,24){$h_3$}}}
\put(250,16){\vector(1,-1){0}}
\put(220,16){\line(1,-1){30}}\put(250,-14){\vector(1,0){40}}
\put(270,-30){\vector(1,0){20}}
\put(300,-22){\makebox(0,0){\framebox(20,24){$h_2$}}}
\put(270,0){\vector(1,0){20}}\put(270,16){\vector(1,0){20}}
\put(310,-30){\vector(1,0){20}}\put(310,-14){\vector(1,0){20}}

\end{picture} \end{center} 
\caption{Changing the order of gates (Lemma~\protect\ref{composition} (b)).} 
\label{fig4} 
\end{figure}
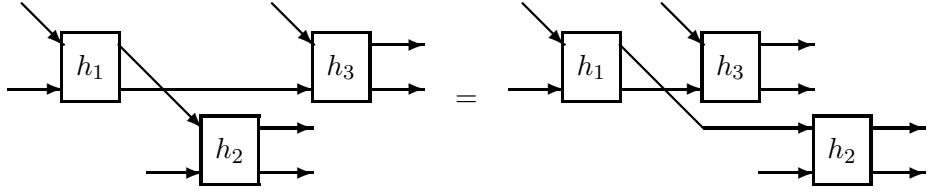

Part (b) follows simply from part (a); as in Figure~\ref{fig4}, part (a) can be 
applied on subcircuit consisting of gates $h_2$ and $h_3$. Note that in this 
case also input qubits $r_j$ of $g_j$'s may be in an entangled state. 
Again a linearity argument shows that we 
have to consider only the case that $r_j$'s are in a product state. \qed

\vspace{5mm}
The above lemma is special case of a more general fact that operations 
on one part of a bi--partite quantum system do not affect the result of operations
on the other part (for more details see, e.g., \cite{nielsen}).

\section{A new equivalent definition for quantum formulas}
\label{equi-section}

Kitaev \cite{kitaev2} has brought to our attention that quantum formulas are 
equivalent to a model that is very similar to the classical formulas.
In this model the inputs and the intermediate results are density matrices.
Each gate is a completely positive trace--preserving super--operator, which 
maps density matrices of a $d$--qubit systems to one--qubit density matrices.
The underlying graph, like a classical formula, is a directed tree; i.e., from
each input there is a unique path to the output gate. Thus the output of such
circuit is a density matrix of a single qubit which provides the probability
of the output ``0'' or ``1''. To make the paper self--contained, we 
first present the definitions of the notions mentioned in this new definition.

By a {\em pure state}\/ $\ket{\alpha}$ we mean a unit vector in some Hilbert space 
$\cc^{2^n}$. A {\em mixed state}\/ $\{\psi\}$ in $\cc^{2^n}$ is a probability 
distribution on pure states in this Hilbert space. We denote such a mixed state as
$\{\psi\}=\left\{ p_k,\ket{\psi_k}\right\}$, where $p_k\geq0$ and $\sum_k p_k=1$. 
Then $\{\psi\}$ picks the pure state $\ket{\psi_k}$ with probability $p_k$.

The {\em density matrix}\/ of a pure state $\ket{\alpha}$ is the 
matrix $\rho_{\ket{\alpha}}$ of the 
linear mapping $\density{\alpha}{\alpha}$; i.e, the mapping 
$\ket{x}\longrightarrow\braket{\alpha}{x}\ket{\alpha}$. So, if 
$\ket{0},\ket{1},\ldots,\ket{2^n-1}$ represent the standard computational basis of
$\cc^{2^n}$ and $\ket{\alpha}=\sum_k\lambda_k\ket{k}$, 
then the $(i,j)$ entry of $\rho_{\ket{\alpha}}$ is $\lambda_i\lambda_j^\star$. 
The importance of density matrix is that it suffices to characterize the quantum
state of the system. Specially, this matrix is enough to find the probabilities
of measurements. In general, the result of each measurement can be represented
by action of a projection operator $\cal P$ on the given state $\ket{\alpha}$,
where $\cal P$ is a projection onto some subspace $\cal E$.
Then the probability that the result of the measurement is in the 
subspace $\cal E$ is equal to 
$\mathrm{Tr}({\cal P}\,\raise.4ex\hbox{$\rho_{\ket{\alpha}}$})$.

The density matrix of a mixed state
$\{\psi\}=\left\{ p_k,\ket{\psi_k}\right\}$ is defined as 
\[ \rho_{\{\psi\}}=\sum_k p_k\rho_{\ket{\psi_k}}
   =\sum_kp_k\density{\psi_k}{\psi_k}. \]
Like the case of pure states, the 
probability that the result of the measurement is in the subspace $\cal E$
is equal to $\mathrm{Tr}({\cal P}\,\raise.4ex\hbox{$\rho_{\{\psi\}}$})$.

If the (pure or mixed) state $\ket{\psi}$ can be written as the tensor
product $\ket{\phi}\otimes\ket{\chi}$ then the density matrix 
$\rho_{\ket{\psi}}$ is equal to the tensor (Hadamard) product 
$\rho_{\ket{\phi}}\otimes\rho_{\ket{\chi}}$.

The next important notion is {\em partial trace}. Consider the Hilbert spaces
${\cal H}_1= \cc^{2^n}$ and ${\cal H}_1= \cc^{2^m}$ and 
${\cal H}={\cal H}_1\otimes{\cal H}_2$; so $\cal H$ is isomorphic with 
$\cc^{2^{n+m}}$. Let 
\[ {\cal B}_1=\left\{\,\ket{u_i}\colon i=1,\ldots,2^n\,\right\} 
   \quad\mbox{and}\quad 
   {\cal B}_2=\left\{\,\ket{v_j}\colon j=1,\ldots,2^m\,\right\} \] 
be orthonormal bases for ${\cal H}_1$ and ${\cal H}_2$, respectively. Then
\[{\cal B}_1\otimes{\cal B}_2=
\left\{\,\ket{u_i}\otimes\ket{v_j}\colon i=1,\ldots,2^n,\ j=1,\ldots,2^m\,\right\}\]
is a basis for $\cal H$. Let $\rho$ be the density matrix of a mixed state 
$\ket{\psi}$ in the space $\cal H$. It is possible to restrict the state 
$\ket{\psi}$ to the subspace ${\cal H}_1$. The result is a {\em partial trace}\/
$\rho\vert_{{\cal H}_1}=\mbox{Tr}_{{\cal H}_2}\,\raise.3ex\hbox{$\rho$}$ 
which is density matrix of some mixed state in the subspace
${\cal H}_1$. We also say that the subspace ${\cal H}_2$ is {\em traced out}.
The partial trace 
$\rho\vert_{{\cal H}_1}$ enables us to calculate probabilities
of the results of the measurements bearing only on the subspace ${\cal H}_1$.
We assume that the rows and columns of the matrices $\rho$ and 
$\rho\vert_{{\cal H}_1}$ are labeled by the vectors in the basis  
${\cal B}_1\otimes{\cal B}_2$ and ${\cal B}_1$, respectively. For example,
$\rho\left(\ket{u_{i_1}}\ket{v_{j_1}},\ket{u_{i_2}}\ket{v_{j_2}}\right)$ 
is the entry of $\rho$ at row labeled by $\ket{u_{i_1}}\otimes\ket{v_{j_1}}$
and the column labeled by $\ket{u_{i_2}}\otimes\ket{v_{j_2}}$. With this
notation, the partial trace $\rho\vert_{{\cal H}_1}$ is defined as follows
\[ \rho\vert_{{\cal H}_1}\left(\ket{u_{i_1}},\ket{u_{i_2}}\right) =\sum_{j=1}^{2^m}
  \rho\left(\ket{u_{i_1}}\ket{v_j},\ket{u_{i_2}}\ket{v_j}\right).\]
Again let $\cal E$ be a subspace of ${\cal H}_1$. We can identify it
with subspace ${\cal E}\otimes{\cal H}_2$ of $\cal H$.
Let ${\cal P}\colon{\cal H}_1\longrightarrow {\cal E}$ be the projection
operator associated with $\cal E$. The operator $\cal P$ can be extended
to the whole space $\cal H$ in a natural way as the operator
${\cal P}\otimes\mbox{Id}_{{\cal H}_2}$, where $\mbox{Id}_{{\cal H}_2}$ is the
identity operator on ${\cal H}_2$.
Then the probability that the result of the measurement is in the 
subspace $\cal E$ is equal to 
$\mbox{Tr}({\cal P}\,\raise.5ex\hbox{$\rho\vert_{{\cal H}_1}$})$.
(For more details on density matrices of mixed states and the partial trace see,
e.g., \cite{cohen}.)

Let $C$ be a quantum circuit. For inputs of $C$ it is possible to consider 
mixed states along with pure states. Toward this end, each input is substituted 
by its density matrix, and each gate $g$ of $C$ by a {\em super--operator}\/
$\widetilde{g}$ that maps density matrices to density matrices. In fact,
if the unitary operator of the gate $g$ is $U$, then
the action of $\widetilde{g}$ on the density matrix $\rho$ is as follows:
\begin{equation}
   \widetilde{g}(\rho)=g\circ\rho=U\rho\,\,U^\dagger. 
\label{tilde-g}
\end{equation}

\begin{lemma}[{\rm \cite{aharonov}}]
If the gates $g_1$ and $g_2$ operate on disjoint sets of qubits, then
for any density matrix $\rho$ we have 
$g_1\circ g_2\circ\rho=g_2\circ g_1\circ\rho$.
\label{density}
\end{lemma}

First we show that every quantum formula is equivalent to a circuit based on
this new definition. Let $\cal F$ be a quantum formula on a basis of $d$--bit 
gates. Construct a circuit $\cal C$ from $\cal F$ by the following transformations. 
In each gate $g$, performing the unitary operation $U\in\mbox{\bf U}\left(2^d\right)$,
keep the only output which is connected to the output and substitute the operator
$U$ by the super--operator $[g]=\mbox{Tr}_{\cal H} \circ \widetilde{g}$, where $\cal H$ is 
the $(d-1)$--dimensional subspace spanned by the qubits removed from the output of this 
gate. The fact that the circuit $\cal C$ computes the same function as the formula
$\cal F$ follows from Lemma~\ref{composition}. Thus the underlying graph of the circuit
$\cal C$ is the same as the computation tree of the formula $\cal F$, where the node
corresponded with the gate $g$ computes the super--operator $[g]$.

Let now $\cal C$ be a circuit based on this new definition. We construct a quantum
formula $\cal F$ from $\cal C$ by simply substituting each gate of $\cal C$,
computing the super--operator $T$, by a $(d+2)$--input $(d+2)$--output unitary gate $U$,
only one output of this gate is connected to the next gate and the other
outputs never interact with any other qubit. So $\cal F$ satisfies our original
definition of quantum formula. The only thing remains is to show how we can choose
the unitary operators $U$ such that the formula $\cal F$ computes the same Boolean
function as $\cal C$. The following theorem guarantees the existence of the correct
operator $U$, for each gate of $\cal C$. Here $\mbox{\bf L}({\cal H})$ is the
space of linear operators on the Hilbert space $\cal H$ and for unitary operator
$U$ on $\cal H$, the operator ${\cal O}_U\in\mbox{\bf L}({\cal H})$ is defined
as ${\cal O}_U(M)= U\, M\, U^\dagger$.

\begin{theorem} [{\rm \cite{choi,kitaev,kraus,schumacher}}]
Suppose that 
$T:\mbox{\bf L}\left({\cal H}_1\right)\longrightarrow\mbox{\bf L}\left({\cal H}_2\right)$
is a trace--preserving and completely positive super--operator. Then there are Hilbert
spaces ${\cal G}_1$ and ${\cal G}_2$, where 
$\dim\left({\cal G}_1\right)=\left(\dim\left({\cal H}_2\right)\right)^2$ and
$\dim\left({\cal G}_2\right)=\dim\left({\cal H}_1\right)\cdot\dim\left({\cal H}_2\right)$,
and there is a unitary operator 
$U:{\cal H}_1\otimes{\cal G}_1\longrightarrow{\cal H}_2\otimes{\cal G}_2$ such that
$T=\mathrm{Tr}_{{\cal G}_2}\circ{\cal O}_U$.
\label{super-operator-theo}
\end{theorem}

We would like to mention that from now on it might be more useful to accept the
new modified definition as the standard one for quantum formulas in the literature.

\section{The lower bound} 
\label{lowerbound}

Let $f(x_1,\ldots ,x_n)$ be a Boolean function. Let
$X=\{x_1,\ldots,x_n\}$ be the set of the input variables. Consider a partition
$\{S_1,\ldots,S_k\}$ of $X$; i.e.,
\[  X=\bigcup_{j=1}^k S_j   \quad\mbox{and}\quad
    S_{j_1}\cap S_{j_2}=\emptyset, \quad\mbox{for}\quad j_1\neq j_2. \]
Let $n_j=|S_j|$, for
$j=1,\ldots,k$. Let ${\cal F}_j$ be the set of all subfunctions of $f$ on
$S_j$ obtained by fixing the variables outside $S_j$ in all possible
ways. We denote the cardinality of ${\cal F}_j$ by $\sigma_j$. 

As an example, we compute the above parameters for the Element Distinctness
function $\mbox{ED}_n$ (see \cite{boppana}). 
Let $n=2\ell\log\ell$ (so $\ell=\Omega(n/\log
n)$) and divide the $n$ inputs of the function into $\ell$ strings each
of $2\log\ell$ bits. Then the value of $\mbox{ED}_n$ is 1 if and only if these
$\ell$ strings are pair wise distinct. We consider the partition 
$(S_1,\ldots,S_\ell)$ such that each $S_j$ contains all variables of the
same string. Thus $n_j=|S_j|=2\log\ell$. Each string in $S_j$ represents an
integer from the set $\{0,1,\ldots,\ell^2-1\}$. The function $\mbox{ED}_n$ is
symmetric with respect to $S_j$'s; so $|{\cal F}_j|=|{\cal F}_{j'}|$. To estimate
$|{\cal F}_1|$, note that if the strings $(z_2,\ldots,z_\ell)$
in $S_2,\ldots,S_\ell$ represent distinct integers then the corresponding 
subfunction is different from any subfunction corresponding to any other string.
So $\sigma_j=|{\cal F}_1|\geq\binom{\ell^2}{\ell-1} > \ell^{\ell-1}$.

\begin{theorem} 
Every quantum formula computing $f$ has size 
\[ \Omega\biggl( 
\sum_{1\leq j\leq k}\frac{\log(\sigma_j)}{\log\log(\sigma_j)}\biggr ). \] 
\label{bound} 
\end{theorem} 

{\bf Proof.} 
We give a proof for any basis consisting of 2--input 2--output quantum 
gates. The proof for bases with more than two inputs is a simple generalization of 
this proof.

Let $F$ be a formula computing $f$. Let $\Sigma_j$ be the
set of input wires of $F$ labeled by a variable from $S_j$, and let
$s_j=|\Sigma_j|$. Then 
\begin{equation} 
 \mathsf{size}(F)=\Omega\biggl( \sum_{1\leq j\leq k}s_j\biggr ) . 
\label{size} 
\end{equation} 
We want
to consider the formulas obtained from $F$ by letting the input
variables not in $\Sigma_j$ to some constant value $\ket{0}$ or $\ket{1}$.
In this regard, let $P_j$ be the set of all paths from an input wire in
$\Sigma_j$ to the output of $F$. Finally, let $G_j$ be the set of gates
of $F$ where two paths from $P_j$ intersect. Then $|G_j|\leq s_j$. 

Let $\tau$ be an assignment of $\ket{0}$ or $\ket{1}$ to the input variable
wires {\em not}\/ in $\Sigma_j$. We denote the resulting formula by
$F_\tau$. Thus $F_\tau$ computes a Boolean function
$f_\tau\colon\{0,1\}^{n_j}\longrightarrow\{0,1\}$ which is a subfunction
of $f$ and a member of ${\cal F}_j$. Consider a path 
\begin{equation}
 \pi=(g_1,g_2,\ldots,g_m), \qquad m > 2, 
\label{path} 
\end{equation} 
in $F_\tau$, where $g_1$ is an input wire or a gate in $G_j$, $g_m$ is a
gate in $G_j$ or the output wire of $F$, and $g_\ell\not\in G_j$ for
$1<\ell<m$.

\begin{figure} \thicklines\begin{center} \unitlength=.5mm 
\begin{picture}(150,52)(0,-30)
\put(0,0){\vector(1,0){30}}\put(0,16){\vector(1,0){30}}
\put(40,8){\makebox(0,0){\framebox(20,24){$\gamma_\ell$}}}
\put(10,5){\makebox(0,0){$q_2$}}\put(10,21){\makebox(0,0){$q_1$}}
\put(50,16){\vector(1,0){30}}
\put(90,8){\makebox(0,0){\framebox(20,24){$\gamma_{\ell+1}$}}}
\put(50,0){\vector(1,-1){30}}\put(80,0){\line(-1,-1){30}}
\put(100,16){\vector(1,0){30}}
\put(140,8){\makebox(0,0){\framebox(20,24){$\gamma_{\ell+2}$}}}
\put(100,0){\vector(1,-1){30}}\put(130,0){\line(-1,-1){30}}
\put(80,0){\vector(1,1){0}}\put(130,0){\vector(1,1){0}}
\put(102,-19){\makebox(0,0){$q_3$}}

\end{picture} 
\end{center}
\caption{The qubits $q_1$ and $q_2$ are strong companions at
step $\ell$, the qubits $q_2$ and $q_3$ are companions at step $\ell+2$.}
\label{fig5} 
\end{figure}
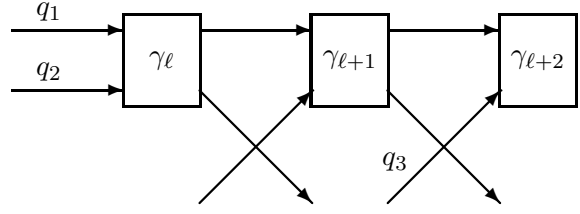

To show how we can squeeze paths like (\ref{path}) (this is the essence
of the Nechiporuk's method), we introduce the following notations. We
consider a natural ordering $\gamma_1,\gamma_2,\ldots,\gamma_t$ on the
gates of the formula $F_\tau$, and regard $F_\tau$ as a computation in
$t$ steps where at step $\ell$ the corresponding gate $\gamma_\ell$ is
performed. We say two qubits $q_1$ and $q_2$ are {\em strong companions}\/
of each other at step $\ell$ if there is a gate $\gamma_j$ such that
$j\leq \ell$ and $q_1$ and $q_2$ are inputs of $\gamma_j$. We say qubits
$q_1$ and $q_2$ are {\em companions}\/ of each other at step $\ell$ if
there exists a sequence $r_1,r_2,\ldots,r_p$ of qubits such that
$r_1=q_1$, $r_p=q_2$, and $r_j$ and $r_{j+1}$ (for $1\leq j\leq p-1$)
are strong companions of each other at step $\ell$ (see Figure~\ref{fig5}). 
If $q_1$ and $q_2$
are companions at step $\ell$ then they are also companions at any step
after $\ell$. For a gate $g=\gamma_k$, we define the {\em set of companions}\/ of 
$g$ as the union of all companions of input qubits of $g$ at step $k$.

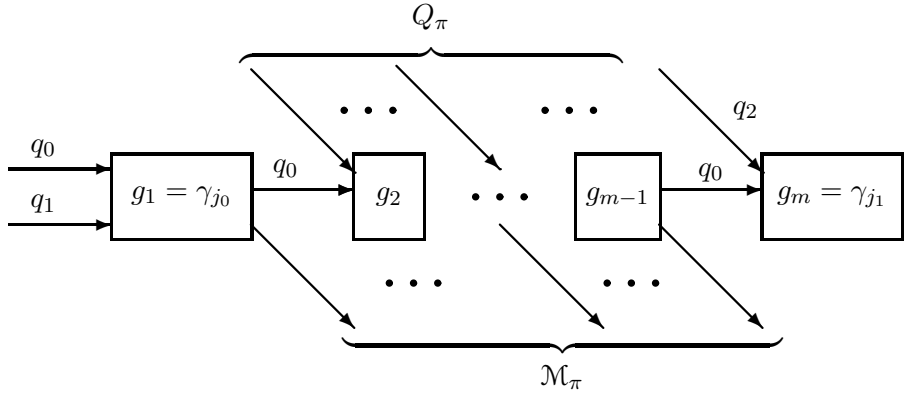
\begin{figure} 
\thicklines
\begin{center} 
\unitlength=.46mm 
\begin{picture}(260,110)(-60,-15)
\put(-60,32){\vector(1,0){30}}\put(-60,48){\vector(1,0){30}} 
\put(-50,38){\makebox(0,0){$q_1$}}\put(-50,54){\makebox(0,0){$q_0$}}
\put(20,48){\makebox(0,0){$q_0$}}
\put(-10,40){\makebox(0,0){\framebox(40,24){$g_1=\gamma_{j_0}$}}}
\put(10,42){\vector(1,0){30}} 
\put(10,32){\vector(1,-1){30}}
\put(50,40){\makebox(0,0){\framebox(20,24){$g_2$}}} 
\put(10,42){\vector(1,0){30}} 
\put(75,40){\circle*{1.5}}\put(82,40){\circle*{1.5}}\put(89,40){\circle*{1.5}}
\put(116,40){\makebox(0,0){\framebox(24,24){$g_{m-1}$}}}
\put(178,40){\makebox(0,0){\framebox(40,24){$g_m=\gamma_{j_1}$}}}
\put(128,42){\vector(1,0){30}}\put(143,47){\makebox(0,0){$q_0$}} 
\put(82,32){\vector(1,-1){30}}\put(128,32){\vector(1,-1){30}}
\put(50,15){\circle*{1.5}}\put(57,15){\circle*{1.5}}\put(64,15){\circle*{1.5}}
\put(113,15){\circle*{1.5}}\put(120,15){\circle*{1.5}}\put(127,15){\circle*{1.5}}
\put(40,47){\line(-1,1){30}}\put(158,47){\line(-1,1){30}}
\put(82,48){\line(-1,1){30}}\put(40,47){\vector(1,-1){0}}
\put(158,47){\vector(1,-1){0}}\put(82,48){\vector(1,-1){0}}
\put(153,65){\makebox(0,0){$q_2$}}
\put(95,65){\circle*{1.5}}\put(102,65){\circle*{1.5}}\put(109,65){\circle*{1.5}}
\put(37,65){\circle*{1.5}}\put(44,65){\circle*{1.5}}\put(51,65){\circle*{1.5}}
\put(100,-12){\makebox(0,0){${\cal M}_\pi$}}
\put(62,92){\makebox(0,0){$Q_\pi$}}
\put(62,81){\makebox(0,0){$\makebox[2in]{\downbracefill}$}}
\put(100,-3){\makebox(0,0){$\makebox[2.3in]{\upbracefill}$}}

\end{picture} 
\end{center} 
\caption{Squeezing a path.}
\label{fig6}
\end{figure}

Suppose that in the path~(\ref{path}) $g_1=\gamma_{j_0}$,
$g_m=\gamma_{j_1}$, the inputs of $g_1$ are $q_0$ and $q_1$,
the output of $\gamma_{j_0}$ from the path~(\ref{path}) 
is the qubit $q_0$, and the input of $\gamma_{j_1}$ not from
the path~(\ref{path}) is the qubit $q_2$ (see Figure~\ref{fig6}). 
Note that $q_0$ is the companion of $q_2$ at step $j_1$. 
Let $Q_\pi$ be the union of all sets of companions of $g_2,\ldots,g_{m-1}$
minus $q_0$ and $q_1$ and their companions at step $j_1$. 
Let $C_0$ be the circuit defined by the gates $g_1,\ldots,g_{m-1}$ from the 
path~(\ref{path}). Suppose that $|Q_\pi|=v$ and consider $C_0$ as an operation
acting on ${\cal H}=\cc^2\otimes\cc^2\otimes\cc^{2^v}$. 
To study the action of the subcircuit $C_0$, it is enough to consider the action
of $C_0$ on the computational basis vectors of the space $\cal H$.
Therefore, while the inputs of $C_0$ as a subcircuit of $F_\tau$ are in general
entangled states, we only have to study the action of $C_0$ on the computational
basis vectors which are product states. We label the
inputs $\ket{\alpha_0}\otimes\ket{\alpha_1}\otimes\ket{\alpha}\in{\cal H}$
of $C_0$ in such a way that when $C_0$ acts as a subformula of $F_\tau$ 
then $\ket{\alpha_0}$, $\ket{\alpha_1}$, and $\ket{\alpha}$ are replaced by
$q_0$, $q_1$, and the companion qubits in $Q_\pi$,
respectively. Note that, because $F_\tau$ is a formula, all qubits in $Q_\pi$ 
are constant inputs of
$F_\tau$ and do not intersect any other path like (\ref{path}).  
So, when $C_0$ acts as a subformula of $F_\tau$,
the input $\ket{\alpha}$ of the subcircuit $C_0$ is 
the same for all possible inputs for
$\ket{\alpha_0}$ and $\ket{\alpha_1}$. Therefore, let 
$\widetilde{C_0}$ be a circuit such that on input
$\ket{\alpha_0}\otimes\ket{\alpha_1}\otimes\ket{0\cdots0}\in
 \cc^2\otimes\cc^2\otimes\cc^{2^v}$, 
it first computes $\ket{\alpha_0}\otimes\ket{\alpha_1}\otimes\ket{\alpha}$ 
then performs the action of $C_0$ on 
$\ket{\alpha_0}\otimes\ket{\alpha_1}\otimes\ket{\alpha}$. 
Then if we replace $C_0$ by $\widetilde{C_0}$ and assign the value
$\ket{0}$ to the qubits in $Q_\pi$, then the result is a circuit equivalent to
$F_\tau$. Suppose that the act of $\widetilde{C_0}$ be defined as follows 
\begin{equation}
  \ket{\alpha_0}\otimes\ket{\alpha_1}\otimes\ket{0\cdots 0}\longrightarrow 
  \sum_{c_0,c_1\in\{0,1\}}\ket{c_0}\otimes\ket{c_1}\otimes
  \ket{A_{c_0,c_1}^{\alpha_0,\alpha_1}} , 
\label{unitary1} 
\end{equation}
where $\alpha_0,\alpha_1\in\{0,1\}$, and
$\ket{A_{c_0,c_1}^{\alpha_0,\alpha_1}}\in\cc^{2^v}$ may be not a unit
vector. Let ${\cal A}_\pi\subseteq \cc^{2^v}$ be the vector space spanned by
$\ket{A_{c_0,c_1}^{\alpha_0,\alpha_1}}$, 
for $\alpha_0,\alpha_1,c_0,c_1\in\{0,1\}$ and $d=\dim({\cal A}_\pi)$. Then
$1\leq d\leq 16$. Let $\ket{A_1^\pi},\ldots,\ket{A_d^\pi}$ be an orthonormal
basis for ${\cal A}_\pi$. Then we can rewrite (\ref{unitary1}) as follows
\begin{equation}
  \ket{\alpha_0}\otimes\ket{\alpha_1}\otimes\ket{0\cdots0}\longrightarrow 
  \sum_{c_0,c_1\in\{0,1\}}\sum_{1\leq j\leq d}
  \lambda_{j,c_0,c_1}^{\alpha_0,\alpha_1}
  \ket{c_0}\otimes\ket{c_1}\otimes\ket{A_j^\pi} . 
\label{unitary2}
\end{equation}

Let ${\cal M}_\pi$ be the set of those unitary operations that are 
performed after one of the gates $g_1,\ldots,g_{m-1}$ on some qubits in $Q_\pi$
before the step $j_1$. Since qubits in $Q_\pi$ do not interact with any 
other path of the form (\ref{path}), by Lemma~\ref{composition} (b), we can postpone
all operations in ${\cal M}_\pi$ after we computed the output
of $g_m$. Let $\pi_1,\ldots,\pi_k$ be a natural ordering on the paths 
like (\ref{path}) on all paths in $P_j$
(i.e., the last gate of $\pi_{j+1}$ is not performed before the last gate of $\pi_j$).
Consider the sets of postponed operations ${\cal M}_{\pi_1},\ldots,{\cal M}_{\pi_k}$.
Once again Lemma~\ref{composition} implies that we can postpone operations in
${\cal M}_{\pi_1}$ after the last gate of $\pi_2$;
then we can postpone operations in ${\cal M}_{\pi_1}$ and ${\cal M}_{\pi_2}$
after the last gate of $\pi_3$, and so on. Repeating this argument
shows that we can postpone all operations in ${\cal M}_{\pi_1},\ldots,{\cal M}_{\pi_k}$
after we compute the output qubit. In this way, the state of the output qubit, before
the postponed operations ${\cal M}_{\pi_1},\ldots,{\cal M}_{\pi_k}$ are applied,
is of the form
\begin{equation} 
  \ket{0}\otimes\ket{M}+\ket{1}\otimes\ket{N} ,
\label{unitary3} 
\end{equation} 
where the first qubit is the output
qubit and $\ket{M}$ and $\ket{N}$ are superpositions of tensor
products of orthonormal vectors $\ket{A_k^{\pi_j}}$ used in (\ref{unitary2}). By
Fact~\ref{basis}, these tensor products of the vectors $\ket{A_k^{\pi_j}}$ are
unit vectors and pair wise orthogonal. The unitary operations in the sets
${\cal M}_{\pi_j}$ (for paths $\pi_j$ of the form (\ref{path})), which are postponed
to the end, do not change the lengths of $\ket{M}$ and $\ket{N}$. Thus,
as far as the computation of the Boolean function $f_\tau$ is concerned,
we can ignore all the postponed unitary operations. For this reason we
construct the circuit $\ov{F_\tau}$ from the formula $F_\tau$ by
eliminating all postponed operations in ${\cal M}_{\pi_j}$, substituting for 
each path $\pi_j$ of the form (\ref{path}) the companion qubits in $Q_{\pi_j}$ 
by four new qubits, and the
unitary operation (\ref{unitary2}) by the operation defined as
\begin{equation}
  \ket{\alpha_0}\otimes\ket{\alpha_1}\otimes\ket{0000}\longrightarrow 
  \sum_{c_0,c_1\in\{0,1\}}\sum_{0\leq j\leq 15}
  \lambda_{j,c_0,c_1}^{\alpha_0,\alpha_1}
  \ket{c_0}\otimes\ket{c_1}\otimes\ket{j} . 
\label{unitary4}
\end{equation}
The output of the circuit $\ov{F_\tau}$, instead of (\ref{unitary3}), is of the form 
\begin{equation}
  \ket{0}\otimes\ket{M'}+\ket{1}\otimes\ket{N'} , 
\label{unitary5}
\end{equation} 
where $\norm{\ket{M}}=\norm{\ket{M'}}$ and
$\norm{\ket{N}}=\norm{\ket{N'}}$. So the circuit $\ov{F_\tau}$ computes
$f_\tau$. Moreover, 
\[ \mathsf{size}(\ov{F_\tau}) =O(s_j), \] 
and for another assignment $\tau'$, the corresponding circuit 
$\ov{F_{\tau'}}$ differs from $\ov{F_\tau}$ only at unitary operations 
defined by (\ref{unitary4}). 

The above discussion implies that $\sigma_j$, the number of subfunctions
on $S_j$, is at most the number of different Boolean functions computed
by size $O(s_j)$ quantum circuits. Therefore, by Theorem~\ref{knill}, we
get \[ \sigma_j\leq 2^{O(s_j\log s_j)} .\] So
$s_j=\Omega({\log(\sigma_j)}/{\log\log(\sigma_j)})$. Now the theorem
follows from (\ref{size}). \qed

\vspace{5mm}
We would like to mention that the fact that a path like (\ref{path})
can be squeezed to a path of constant length is a special case of the
general property of super--operators stated in Theorem~\ref{super-operator-theo}.

To apply the general bound of the above theorem, we could
consider any of the several explicit functions used in the case of
Boolean formulas (see \cite{dunne,wegener}). As we mentioned in 
the beginning of this section, we consider the Element Distinctness function
$\mbox{ED}_n$. For this function
$\sigma_j>\ell^{\ell-1}$, where $\ell=\Omega(n/\log n)$. Therefore, we get 
the lower bound $\Omega(\ell^2)=\Omega(n^2/\log^2 n)$ for the formula size. 

\begin{theorem} 
Any quantum formula computing $\mbox{\em ED}_n$ has size
$\Omega(n^2/\log^2 n)$.
\end{theorem}

\section{Quantum formulas vs. Boolean circuits}

In this section we show that quantum formulas are not more powerful than
Boolean circuits. So as a model of computation, their strength lies between
Boolean formulas and Boolean circuits.

Following the idea developed in Section~\ref{equi-section}, we consider 
a quantum formula as a quantum circuit operating on {\em mixed states}.
For the details of quantum circuits with mixed states see \cite{aharonov,kitaev}.
Before we start the proof of the main result of this section, 
we need to see how we can bound errors in quantum circuits with mixed states.
Toward this end we need a suitable norm on super--operators. Each
super--operator $T$ which maps density matrices to density matrices is
a linear mapping  of the form 
$\mbox{\bf L}({\cal H}_1)\longrightarrow\mbox{\bf L}({\cal H}_2)$,
where ${\cal H}_1$ and ${\cal H}_2$ are finite--dimensional Hilbert spaces and
$\mbox{\bf L}({\cal H}_j)$ is the set of linear operators on ${\cal H}_j$.
Note that $\mbox{\bf L}({\cal H}_j)$ itself is a linear space.
Let $\cal H$ be an $m$--dimensional Hilbert space.
There are several norms on the space $\mbox{\bf L}({\cal H})$, of which
we need the following ones. Let $A\in\mbox{\bf L}({\cal H})$. We
identify $A$ with its $m\times m$ matrix $(a_{ij})$. 
The first norm is 
\[ M(A)=m\max_{i,j}\lvert a_{ij}\rvert .\]
The usual norm is defined as 
\[ \norm{A}=\sup_{\ket{x}\neq0}\frac{\norm{A\ket{x}}}{\norm{\ket{x}}} =
    \max\left\{\,\sqrt{\lambda} \colon \lambda\in\mbox{Spec}(A^\dagger A)  
       \,\right\}, \]
where $\mbox{Spec}(M)$ is the {\em spectrum}\/ of the matrix $M$; i.e., the set
of the eigenvalues of $M$. The other norm is the {\em trace norm}:
\[ \norm{A}_{\mathrm{Tr}}=\sum_{\lambda\in{\mathrm{Spec}}(A^\dagger A)}
   \sqrt{\lambda}. \]
We need the next norm $\norm{\cdot}_\star$ to define another norm: let $T$ be a linear
operator that maps matrices to matrices; i.e., 
$T\in\mbox{\bf L}(\mbox{\bf L}({\cal H}))$, then
\[ \norm{T}_\star=\sup_{A \neq 0}
   \frac{\norm{TA}_{\mathrm{Tr}}}{\norm{A}_{\mathrm{Tr}}} .\]
The last norm we consider is the {\em diamond norm}, defined in 
\cite{kitaev} and also in \cite{aharonov}. 
To define this norm, we consider a Hilbert space $\cal G$ such that
$\dim({\cal G})\geq\dim({\cal H})$ and we let
\[ \norm{T}_\diamond=\norm{T\otimes I_{\cal G}}_\star, \] 
where $I_{\cal G}$ is the identity operator on $\cal G$.
The followings are the basic properties of these norms.

\begin{itemize}

\item[(i)] $\frac{1}{m}M(A)\leq\norm{A}\leq M(A)$.

\item[(ii)] $\norm{A}_{\mathrm{Tr}}\leq m\norm{A}$.

\item[(iii)] $\norm{T(\rho)}_{\mathrm{Tr}}\leq\norm{T}_\diamond\norm{\rho}_{\mathrm{Tr}}$, 
           for the density matrix $\rho$.

\item[(iv)] $\norm{TR}_\diamond\leq\norm{T}_\diamond\norm{R}_\diamond$.

\item[(v)] $\norm{T\otimes R}_\diamond=\norm{T}_\diamond\norm{R}_\diamond$.

\item[(vi)] If $T=\widetilde{g}$, for some quantum gate $g$, or 
           $T={\mathrm{Tr}}_{\cal F}$, then $\norm{T}_\diamond=1$.
           

\end{itemize}
The properties (iii)--(vi) are proved in \cite{aharonov,kitaev}.

For any operator $V\in\mbox{\bf L}({\cal H})$ we define
the operator ${\cal O}_V\in\mbox{\bf L}(\mbox{\bf L}({\cal H}))$ as
\begin{equation}
   {\cal O}_V(M)=V\,M\,V^\dagger,\qquad M\in\mbox{\bf L}({\cal H}).
\label{cal-o}
\end{equation}
In \cite{aharonov,kitaev} it is proved that 
$\norm{{\cal O}_V-{\cal O}_W}_\diamond\leq 2\norm{V-W}$
if $\norm{V}\leq1$ and $\norm{W}\leq1$. We need the following general
form of this inequality.

\begin{lemma}
Let $\dim({\cal H})=m$.
For any $V,W\in\mbox{\bf L}({\cal H})$ we have
\[ \norm{{\cal O}_V-{\cal O}_W}_\diamond\leq 2m\,\norm{V-W}\,\min(\norm{V},\norm{W})
 +m\norm{V-W}^2. \]
\label{ov-ow}
\end{lemma}

{\bf Proof.} 
We have (for $A\in\mbox{\bf L}\left({\cal H}\otimes{\cal H}\right)$)
\begin{eqnarray}
\norm{{\cal O}_V-{\cal O}_W}_\diamond & = & \sup_{A\neq0}
   \norm{({\cal O}_V\otimes I_{\cal H})A-({\cal O}_W\otimes 
   I_{\cal H})A}_{\mathrm{Tr}} / \norm{A}_{\mathrm{Tr}} \nonumber \\
 & = & \sup_{A\neq0}{\norm{(V\otimes I_{\cal H})\,A\,
  (V^\dagger\otimes I_{\cal H})-(W\otimes I_{\cal H})\,A\,
   (W^\dagger\otimes I_{\cal H})}_{\mathrm{Tr}}} / {\norm{A}_{\mathrm{Tr}}} \nonumber \\
 & = & \sup_{A\neq0}\Vert(V\otimes I_{\cal H})\,A\,
  (V^\dagger\otimes I_{\cal H})- \nonumber \\
 &   & ((V+(W-V))\otimes I_{\cal H})\,A\,
   ((V^\dagger+(W^\dagger-V^\dagger))\otimes 
   I_{\cal H})\Vert_{\mathrm{Tr}} / \norm{A}_{\mathrm{Tr}} \nonumber \\
 & \leq & \sup_{A\neq0}\norm{(V\otimes I_{\cal H})\,A\,((W^\dagger-V^\dagger)
     \otimes I_{\cal H})}_{\mathrm{Tr}}/\norm{A}_{\mathrm{Tr}} + \nonumber \\
 &      & \sup_{A\neq0}\norm{((W-V)\otimes I_{\cal H})\,A\,
     (V^\dagger\otimes I_{\cal H})}_{\mathrm{Tr}}/\norm{A}_{\mathrm{Tr}} + \label{eq1} \\
 &      & \sup_{A\neq0}\norm{((W-V)\otimes I_{\cal H})\,A\,
    ((W^\dagger-V^\dagger)\otimes I_{\cal H})}_{\mathrm{Tr}}/
    \norm{A}_{\mathrm{Tr}}. \nonumber
\end{eqnarray}
Since $\norm{MN} \leq \norm{M} \cdot \norm{N}$,
$\norm{M}\leq\norm{M}_{\mathrm{Tr}}\leq m\norm{M}$, 
$\norm{M\otimes N}=\norm{M}\cdot\norm{N}$, 
$\norm{M^\dagger}=\norm{M}$, and $\norm{I_{\cal H}}=1$, it follows that
\begin{eqnarray*}
\norm{(V\otimes I_{\cal H})\,A\,((W^\dagger-V^\dagger)
     \otimes I_{\cal H})}_{\mathrm{Tr}}
  & \leq & m \norm{(V\otimes I_{\cal H})\,A\,((W^\dagger-V^\dagger)
     \otimes I_{\cal H})} \\
  & \leq & m \norm{V\otimes I_{\cal H}} \cdot \norm{A} \cdot
          \norm{(W^\dagger-V^\dagger)\otimes I_{\cal H}} \\
  & \leq & m \norm{V} \cdot \norm{A}_{\mathrm{Tr}} \cdot \norm{W-V}.
\end{eqnarray*}
By applying a similar reduction to the other terms of (\ref{eq1}), we drive the
following inequality
\[ \norm{{\cal O}_V-{\cal O}_W}_\diamond\leq 2m\norm{V-W}\cdot\norm{V}
   +m\norm{V-W}^2. \]
We can also drive a similar inequality with $\norm{V}$ substituted by 
$\norm{W}$. This completes the proof. \qed

\vspace{5mm}
We say two $n\times m$ matrices $A=(a_{ij})$ and $B=(b_{ij})$ are 
{\em $\delta$--close}\/ to each other if $\vert a_{ij}-b_{ij}\vert\leq\delta$,
for every $1\leq i\leq n$ and $1\leq j\leq m$. If the $m\times m$ matrices
$A$ and $B$ are $\delta$--close to each other then
\begin{equation}
    \norm{A-B}\leq M(A-B)\leq m\,\delta. 
\label{a-b}
\end{equation}

The following theorem formalizes the general form of the error bound for 
quantum circuits when approximating the unitary operator of each gate.
This theorem is actually a generalization of a weaker theorem which has appeared
in several papers (see, e.g., \cite{aharonov,vazirani,kitaev}).
We need this generalization because once we substitute
any unitary gate $S$ of the original quantum circuit by some approximated gate $T$, 
in general we do not know whether $\norm{T}\leq 1$ or not (this is the assumption 
of the weaker version of this theorem).

\begin{theorem}
Let $S_j,T_j\in\mbox{\bf L}\left(\mbox{\bf L}\left(\cc^{2^d}\right)\right)$, 
$1\leq j\leq\ell$, be defined as $S_j={\cal O}_{U_j}$ and $T_j={\cal O}_{V_j}$,
where $U_j\in\mbox{\bf U}(2^d)$ is unitary and $V_j$ is $\delta$--close
to $U_j$. Then
\[ \norm{S_\ell\cdots S_3S_2S_1-T_\ell\cdots T_3T_2T_1}_\diamond\leq 
    e^{\eta(d,\delta)\ell}-1,\]
where $\eta(d,\delta)=2^{2d+1}\delta\left(1+2^d\delta\right)$.
\label{app-th}
\end{theorem}

{\em Proof.}
First note that
\begin{alignat*}{2}
  \norm{S_j-T_j}_\diamond 
    & =\norm{{\cal O}_{U_j}-{\cal O}_{V_j}}_\diamond &&\\
    & \leq 2^{d+1}\norm{U_j-V_j}\left(1+\norm{U_j-V_j}\right) 
      && \qquad\mbox{by Lemma~\ref{ov-ow}}\\
    & \leq 2^{2d+1}\delta\left(1+2^d\delta\right) && \qquad\mbox{by (\ref{a-b})}\\
    & = \eta(d,\delta); &&
\end{alignat*}
and, by (vi),
\begin{equation}
   \norm{T_j}_\diamond \leq \norm{S_j}_\diamond+
   \norm{S_j-T_j}_\diamond \leq 1+\eta(d,\delta) .
\label{t-norm}
\end{equation}
Also we have the following simple inequality
\begin{eqnarray}
 \norm{M_2M_1-N_2N_1}_\diamond 
  & =    & \norm{M_2(M_1-N_1)-(M_2-N_2)N_1}_\diamond \nonumber \\
  & \leq & \norm{M_2}_\diamond\norm{M_1-N_1}_\diamond+
        \norm{N_1}_\diamond\norm{M_2-N_2}_\diamond \label{mn}
\end{eqnarray}
Now, by repeated applications of (\ref{t-norm}) and (\ref{mn}), we have
\begin{eqnarray*}
 \norm{S_\ell\cdots S_3S_2S_1-T_\ell\cdots T_3T_2T_1}_\diamond
   & \leq & \norm{S_\ell\cdots S_3S_2}_\diamond\norm{S_1-T_1}_\diamond+ \\
   &   & \norm{T_1}_\diamond\norm{S_\ell\cdots S_3S_2-T_\ell\cdots T_3T_2}_\diamond \\
   & \leq & \eta(d,\delta)+(1+\eta(d,\delta))
            \norm{S_\ell\cdots S_3S_2-T_\ell\cdots T_3T_2}_\diamond \\
   & \leq & \eta(d,\delta)+\eta(d,\delta)(1+\eta(d,\delta))+ \\
   &      & (1+\eta(d,\delta))^2
            \norm{S_\ell\cdots S_3-T_\ell\cdots T_3}_\diamond \\
 \vdots &   & \vdots \\
   & \leq & \eta(d,\delta)\sum_{j=0}^{\ell-1}\big(1+\eta(d,\delta)\big)^j \\
   & = & (1+\eta(d,\delta))^\ell-1 \\
   & \leq & e^{\eta(d,\delta)\ell}-1. \qquad \qed                 
\end{eqnarray*}

The following theorem is the immediate consequence of the above theorem.
Note that for a gate $g$ the super--operator $\widetilde{g}$ is defined by 
(\ref{tilde-g}).

\begin{theorem}
Let $C$ be a quantum circuit composed of the gates $g_1,\ldots,g_s$.
Suppose that each $g_j$ is a $d$--bit gate computing the unitary operator
$U_j\in\mbox{\bf U}(2^d)$. For each $1\leq j\leq s$, let 
$V_j\in\mbox{\bf L}\left(\cc^{2^d}\right)$ be a $\delta$--close matrix to $U_j$. 
Let $T_j\in\mbox{\bf L}\left(\mbox{\bf L}\left(\cc^{2^d}\right)\right)$ 
be defined as $T_j={\cal O}_{V_j}$. For any input density matrix $\rho_0$, let 
\[ \psi=(\widetilde{g}_s\otimes I_{{\cal H}_s})\circ\cdots\circ
   (\widetilde{g}_2\otimes I_{{\cal H}_2})\circ(\widetilde{g}_1\otimes 
   I_{{\cal H}_1}) \, \rho_0 \]        
be the output of $C$, where ${\cal H}_j$ is the Hilbert space 
generated by the qubits not involved with the gate $g_j$. Also, let
\[  \zeta=(T_s\otimes I_{{\cal H}_s})\cdots (T_2\otimes I_{{\cal H}_2})
    (T_1\otimes I_{{\cal H}_1}) \, \rho_0 \]
 be the approximated output of $C$. Then 
\[\norm{\psi-\zeta}_{\mathrm{Tr}} \leq \left(e^{\eta(d,\delta)s}-1\right)
     \norm{\rho_0}_{\mathrm{Tr}},\]
where $\eta(d,\delta)=2^{2d+1}\delta\left(1+2^d\delta\right)$.
\label{approx-th}
\end{theorem}

\begin{theorem}
Let $\cal B$ be a quantum basis. Then each quantum formula of size $\ell$ 
and depth $d$ over the basis $\cal B$ can be simulated with error at most 
$\vep$ by a Boolean {\em circuit}\/ of size 
$O\left(\ell\,\mu\,\log\mu\log\log\mu\right)$ and depth 
$O\left(d\,\log\mu\right)$, where $\mu=\lceil\log\ell-\log\vep\rceil$.
\label{simulation}
\end{theorem}

{\bf Proof.} 
The basic idea of the simulation is to look at the behavior
of a quantum formula as a quantum circuit acting on density matrices of 
mixed states. We assume, w.l.o.g.,  that each gate in the basis $\cal B$ is a
2--bit gate.

Consider a quantum formula $\cal F$ over the basis $\cal B$; suppose 
that $\cal F$ has $t$ inputs (constant or variable) and
computes the Boolean function $f\colon\{0,1\}^n\longrightarrow\{0,1\}$.
We show that  there is a Boolean circuit $\cal C$ that for any input
$\bsym{a}=(a_1,\ldots,a_n)\in\{0,1\}^n$ simulates the action of $\cal F$ on $\bsym{a}$.
Let $\ket{\alpha}=\ket{0}\otimes\ket{A_0}+\ket{1}\otimes\ket{A_1}$ be the 
output of $\cal F$ on the input $\bsym{a}$. Suppose that the first qubit is the 
output bit. If we {\em trace out}\/ the non--output
bits of $\ket{\alpha}$, the result is a $2\times 2$ density matrix 
$\rho_{\mathrm{final}}=\rho\vert_{\ket{\alpha}}$. From $\rho_{\mathrm{final}}$ 
it is easy to calculate the probability
of acceptance of $\cal F$. The formula structure of $\cal F$ allows us to calculate the
density matrix $\rho_{\mathrm{final}}$ without going to the $2^t$ dimensional space.
The Boolean circuit $\cal C$ finds the density matrix $\rho_{\mathrm{final}}$
by simulating the gates of $\cal F$ step by step.

Since the trace norm of a density matrix is equal to its trace,
it follows that $\norm{\rho_{0}}_{\mathrm{Tr}}=1$, where $\rho_{0}$ is
the density matrix of the input.

Now the gates of $\cal F$ are no longer acting on pure states,
but they are acting on mixed states. If the input of a gate $g_j$,
(performing the unitary operation $U_j$)
is the $4\times 4$ density matrix $\rho$ then the output is the density
matrix $\rho'=U_j\,\rho\,\,{U_j}^\dagger$. 
Of the two output bits $q_1$ and $q_2$ of this gate 
only one, say $q_1$, is connected to the output bit of $\cal F$.
So we trace out the system
representing $q_2$ and consider the new density matrix 
$\rho\vert_{q_1}=\mbox{Tr}_{q_2}\rho'$ for $q_1$. By repeating this process for each 
gate of $\cal F$ we finally get the desired density matrix $\rho_{\mathrm{final}}$.
The correctness of this process follows from Lemma~\ref{density}.

The Boolean circuit $\cal C$ can simulate the calculations of these density matrices
$\rho_{q_1}$. The only problem for this simulation is the proper approximation of the
entries of unitary matrices $U_j$. If we substitute each entry of $U_j$ by its
first $\mu=-\lceil\log_2\delta\rceil$ bits, then we get a matrix that is  
$\delta$--close to $U_j$. Let $\widetilde{\cal F}$ be the resulting formula and
$\widetilde{\rho}_{\mathrm{final}}$ be the output of $\widetilde{\cal F}$.
Then, by Theorem~\ref{approx-th}, 
$\norm{\rho_{\mathrm{final}}-\widetilde{\rho}_{\mathrm{final}}}_{\mathrm{Tr}}
 \leq e^{\eta(d,\delta)\ell}-1$. 
So if $\delta=O\left(\frac{\vep}{\ell}\right)$, i.e., 
$\mu=O\left(\log\ell-\log\vep\right)$, then the simulation of $\cal F$ by
$\widetilde{\cal F}$ has at most $\vep$ error. The theorem now follows from this
fact that addition and multiplication of $m$ bits numbers can be carried out
by Boolean circuits of size $O(m\log m\log\log m)$ and depth $O(\log m)$ 
(see \cite{schonhage,wegener}). \qed

\vspace{5mm}
Why does this proof not provide a Boolean {\em formula}\/ instead of Boolean
circuit? The reason is that to calculate $\rho'=U_j\,\rho\,\,{U_j}^\dagger$,
we need 4 copies of each entry of $\rho$. Thus the fan--out of the
gates in the Boolean circuit obtained from the formula $\widetilde{\cal F}$ is
4. This means that the Boolean formula equivalent to this Boolean circuit,
in general, has size exponential in $\ell$; this size is at least
$\Omega\left(\ell^3\right)$, if the graph of $\cal F$ is a full binary tree.

\section{Concluding Remarks} 

We have extended a classical technique for proving lower bound for Boolean
formula size to quantum formulas. The difficult part was to effectively deal
with the phenomenon of entanglement of qubits. While we have been successful
in extending a classical technique to the quantum case, the challenges 
encountered indicate that in general the problem of extending methods of Boolean
case to the quantum case may not have simple solutions. 
For example, even the seemingly simple issue of the exact 
relationship between quantum formulas and quantum circuits has not been resolved.
In the Boolean case, simulation of circuits by formulas is a simple 
fact, but in the quantum case it is not clear whether every quantum circuit 
can be simulated by a quantum formula. In particular, it is not clear that in 
the process of going from quantum circuits to formulas, how we can modify 
the underlying entanglement of qubits while keeping the probability of 
reaching to the final answer the same. We were also able to show that it 
is possible to simulate quantum formulas with Boolean circuits of almost the 
same size. It does not seem that Boolean formulas could efficiently simulate 
their quantum counterparts. So evidently quantum formulas, as a model of 
computation, are more powerful than Boolean formulas and less powerful than 
Boolean circuits. A better understanding of the relations between these models 
remains a challenging problem.

\section{Appendix: Counting the number of Boolean functions computed by 
quantum circuits of a given size}

In this appendix we prove the following upper bound.

\begin{theorem} 
The number of different $n$--variable Boolean functions that 
can be computed by size $N$ quantum circuits ($n\leq N$) with $d$--input 
$d$--output elementary gates (for some constant $d$)
is at most $2^{O(nN)+O(N\log N)}$.  
\end{theorem} 

Our proof is based on Warren's bound on the number of different sign--assignments to
real polynomials \cite{warren}. We begin with some necessary notations.

Let $P_1(x_1,\ldots,x_t),\ldots,P_m(x_1,\ldots,x_t)$ be real polynomials. 
A {\em sign--assignment}\/ to these polynomials is a system of inequalities
\begin{equation}
  P_1(x_1,\ldots,x_t)\,\Delta_1\,0, \quad \ldots, \quad 
  P_m(x_1,\ldots,x_t)\,\Delta_m\, 0 , 
\label{sign}
\end{equation}
where each $\Delta_j$ is either ``$<$'' or ``$>$''. The sign--assignment 
(\ref{sign}) is called {\em consistent}\/ if this system has a solution in $\rr^t$.

\begin{theorem} [{\rm Warren \cite{warren}}]
Let $P_1(x_1,\ldots,x_t),\ldots,P_m(x_1,\ldots,x_t)$ be real polynomials, 
each of degree at  most $d$. Then there are at most $(4edm/t)^t$ consistent 
sign--assignments of the form {\em (\ref{sign})}.
\label{warrentheo}
\end{theorem}

We consider the class of quantum circuits 
of size $N$ with $d$--bit gates computing $n$--variable Boolean functions. 
Without loss of generality, we can assume that $n'$, the number of input wires 
of such circuits, is at most $d\cdot N$. We define an equivalence relation
$\simeq$ on such circuits: we write $C_1 \simeq C_2$ if and only if
$C_1$ and $C_2$ differ only in
the label of their gates; in another word, $C_1$ and $C_2$ have the same underlying
graph but the corresponding gates in these circuits may compute different unitary
operations. The number of different equivalence classes is at most
\[ \binom{n'}{d}^{N} \leq (dN)^{dN} = 2^{O(N\log N)}.\]
Now we find an upper bound for the number of different Boolean functions that 
can be computed by circuits in the same equivalence class. 
Fix an equivalence class $\cal E$. We use the variables
$a_1+ib_1,a_2+ib_2,\ldots,a_\mu+ib_\mu$, where $\mu=d^2N$, to denote the 
entries of the matrices of
the gates of a circuit $C$ in $\cal E$. By substituting appropriate values to the
variables $a_1,\ldots,a_\mu,b_1,\ldots,b_\mu$, we get all circuits in $\cal E$.
On input $\alpha=(\alpha_1,\ldots,\alpha_n)\in\{0,1\}^n$, the probability
that $C$ outputs 1 can be represented by a {\em real}\/ polynomial 
\[ P_\alpha(a_1,\ldots,a_\mu,b_1,\ldots,b_\mu). \]
The degree of $P_\alpha$ is 
at most $2N$. There are $2^n$ polynomials $P_\alpha$ and the number of 
different Boolean functions can be computed by $C$ by changing the unitary 
operators of its gates is at most the
number of different consistent sign--assignments to the following system:
\[ \mbox{$P_\alpha(a_1,\ldots,a_\mu,b_1,\ldots,b_\mu)-\frac{2}{3}$}, \]
\[ \mbox{$P_\alpha(a_1,\ldots,a_\mu,b_1,\ldots,b_\mu)-\frac{1}{3}$}, \]
for $\alpha\in\{0,1\}^n$.
By Theorem~\ref{warrentheo} this number is bounded from the above by
\[\biggl ( \frac{4e(2N)2^{n+1}}{2\mu} \biggr )^{2\mu}=
         2^{O(nN)+O(N\log N)} .\qquad \qed  \]

\vspace{5mm}
{\bf Acknowledgments.}
We thank Alexei Kitaev for helpful discussions on the early version of this paper.
We also wish to thank the reviewer for careful reading and useful comments.


\end{document}